\documentclass[aps,twocolumn,showpacs,preprintnumbers,amsmath,amssymb,floatfix]{revtex4}
\usepackage{graphicx}
\usepackage{epsfig}
\begin{document}
\def\be{\begin{equation}}
\def\ee{\end{equation}}
\def\bearr{\begin{eqnarray}}
\def\eearr{\end{eqnarray}}
\def\tc{$T_c~$}
\def\bis2{$\rm BiS_2$~}
\def\spx{$\rm 6p_x$~}
\def\spy{$\rm 6p_y$~}
\def\laob{$\rm LaOBiS_2$~}
\def\laofh{$\rm LaO_{0,5}F_{0.5} BiS_2$~}
\def\laofx{$\rm LaO_{1-x}F_{x} BiS_2$~}
\def\half{$\rm \frac{1}{2}$~}

\title{Two Mott Insulator Theory of Superconductivity in K$_3$X\\(X: picene, .. p-terphenyl, .. C$_{60}$)}
\author{G. Baskaran}

\affiliation
{The Institute of Mathematical Sciences, C.I.T. Campus, Chennai 600 113, India  \&\\
Perimeter Institute for Theoretical Physics, Waterloo, ON N2L 2Y5, Canada}

\begin{abstract} 
We look for unifying aspects behind superconductivity in aromatic hydrocarbon and fullerene family
K$_3$X (X: picene, .. p-terphenyl, .. C$_{60}$). Aromatic hydrocarbon molecules support RVB states. Consequent stability (aromaticity) makes them reluctant electron acceptors. We argue that X accepts only two (not all three) electrons from K$_3$ and creates charged RVB's in X$^{2-}$, and becomes a (molecular) Cooper pair box. A weak Josephson coupling between X$^{2-}$ molecules creates a Bose Mott insulator, a potential high Tc superconductor. Remaining lone electron in the complex (K$_3)^{2+}$ occupies a suitable metal orbital hybrid. They hybridize weakly through X$^{2-}$ molecular bridges, to form a half filled band of renormalized K atom orbitals, a Fermionic Mott insulator. An interplay of RVB physics and charge transfer (mutual doping) or external doping leads to superconductivity in one or both Mott insulators. In our theory there is room for room temperature superconductivity.
\end{abstract}

\maketitle

\section{Introduction}

Organic route to high Tc superconductivity is highly desirable. Ever since a theoretical proposal by Little \cite{Little}, of organic polymer based high Tc superconductivity, the search continues.  Organic solids such as TTF-TCNQ and doped polyacetylene which drew a lot of attention at the beginning have not become superconducting. However, several aromatic hydrocarbon \cite{picene} and C$_{60}$ molecule \cite{C60} based superconductors have been synthesized \cite{CarbonSupCondRev,orgSupCondBook,C60Review}: K$_3$X (X: picene, phenanthrene, dibenzo pentacene, coronene, ... C$_{60}$). Some of them are high Tc superconductors. Other low Tc organic superconductor families, Bechgard salts and ET salts \cite{orgSupCondReview}, continue to excite physics community from the point of view of strong correlation physics.

A recent claim of superconductivity \cite{ptpSupCond,ptpSearch} at an elevated temperature of 123 K in K$_3$(p-terphenyl) is remarkable and exciting. (Three benzene rings bridged by two C-C bonds along a line forms para (p-) terphenyl \cite{picenePterphenyl}, C$_{18}$H$_{14}$). A more recent high resolution photoelectron spectroscopic evidence \cite{ptpPES} for pairing gap in p-terphenyl system upto 60 K adds to the excitement and makes the activity worth pursuing. While Meissner signals and a favourable magnetic field dependence of Tc are evidences for superconductivity, other standard and confirmatory measurements are needed. Are we witnessing birth of a new revolution in high Tc superconductivity ?

Aim of the present article is to suggest an unified model and mechanism of superconductivity in the family of alkali metal doped aromatic hydrocarbon and fullerene molecular solids K$_3$X. 

Briefly, we find that K$_3$X has an unusual nominal charge state $(K_3)^{2+}(X)^{2-}$, rather than usually assumed $(K_3)^{3+}(X)^{3-}$. A Boson Mott insulator (even electron dication subsystem X$^{2-}$) and Fermion Mott insulator (odd electron dianion subsystem $(K_3)^{2+}$) emerge. Charge transfer between two Mott insulators or external doping can create superconductivity in one or both Mott insulators. That is, in our theory both anion and cation subsystems could become active participants in superconductivity. In existing theories \cite{GBTosattiC60,KivelsonC60,PairbindingOtherWorkds,pairBindingKivelson2016,ChinesePairBinding,MicheleErioTheoryC60,
pTheoryCapone,pTheoryKim,pTheoryAoki,pTheoryValenti,pTheoryJoern,pTheoryYamada,pTheoryDuttaMajumdar} of superconductivity, a nominal charge state $(K_3)^{3+}(X)^{3-}$ makes $(K_{3})^{3+}$ effectively an insulating spectator and $(X)^{3-}$ effectively becomes a (odd electron) molecular conductor (with some hybridization with K orbitals), and supports superconductivity based on Mott physics.

Firstly, we recognize that polycyclic aromatic hydrocarbon molecules X are different from molecules with saturated bonds. Molecule X supports a correlated many electron state, a p-$\pi$ pool of neutral singlets (molecular RVB state). In our theory molecule X accepts only two valence electrons, accommodates the pair in valence bond resonance and becomes a (molecular) Cooper pair box X$^{2-}$. Josephson coupling between X$^{2-}$ creates a Boson Mott insulator, a potential high Tc superconductor. Further, the molecular matrix X$^{2-}$ isolates the dication complex (K$_3)^{2+}$. Remaining lone electron in (K$_3)^{2+}$ occupies a suitable metal atom hybrid state. This hybrid further hybridizes weakly with LUMO of bridging molecules X$^{2-}$ and forms a renormalized metal atom Mott insulator. This fermionic Mott insulator becomes a template for high Tc superconductivity via RVB mechanism \cite{PWAScience,BZA}. Mutual charge transfer between two Mott 
insulators or external doping can create superconductivity in one or both Mott insulators.

In building our theory we use some key quantum chemical and experimental facts: i) polycyclic aromatic hydrocarbon molecules have in general low electron affinities \cite{ElectronegativityTable,EAffinityTheory}, due to their extra stability (aromaticity) arising from valence bond resonances. For example, benzene has negative electron affinity $ \approx $ - 1.5 eV. Consequently electron affinities of polycyclic aromatic hydrocarbons (fused benzene rings) are small in general. They are reluctant electron acceptors in the solid state, ii) an ubiquitous stoichiometric composition K$_3$X, (X = p-terphenyl, picene, dibenzopentancene, coronene, C$_{60}$, ...) in all known superconductors and absence of superconductivity for stoichiometry K$_2$ as well as K$_1$, iii) reduced Madelung energy from strong dielectric screening and iv) evidence from X-ray spectroscopy \cite{KMixingXPS} for a substantial orbital hybridization of valence orbitals of K and molecular orbitals of picene at the fermi energy in K$_3$(picene).

Our theory uses earlier and recent theoretical works and insights: i) recognizing the molecule X as a strongly correlated electron system, capable of supporting charged Cooper pair inside molecule X via pair binding and \cite{GBTosattiC60,KivelsonC60} intermolecular Josephson tunnelling \cite{KrasinkovaJosephsonJn} ii) a clue from a recent theoretical calculations by Naghavi and Tosatti \cite{pTheoryErio1,pTheoryErio2} and by Chiappe et al. \cite{pTheoryChiappe}, which indicate transfer of two rather than all three electrons to X in K$_3$X for X = picene and phenanthrene,iii) Mott insulating character for the stoichiometry K$_2$, \cite{pTheoryValenti,pTheoryYamada}, iv) a recognition \cite{pTheoryYamada} that a Mott insulating reference solid K$_2$X could be important for creating superconductivity.

Our paper is organized as follows. We first discuss, using elementary considerations, why two electron rather than three electron transfer is favoured in K$_3$X.  Then we discuss how a charged dianion X$^{2-}$ effectively becomes a Cooper pair box. This is followed by discussion of a weak Josephson coupling between X$^{2-}$ molecules and formation of a Boson Mott insulator, a potential superconductor. A discussion of formation of K atom based Wannier function and emergence of a Fermion Mott insulator is discussed next.  A Hubbard model for the renormalized metal atom Mott insulator is presented. Issues of charge transfer from Boson Mott insulator to Fermion Mott insulator (mutual doping), self doping or external doping and estimation of Tc are discussed next. 

Before the concluding section we discuss applicability of our ideas to alkali metal doped superconducting fullerides, including Mott insulating Cs$_3$C$_{60}$ that has A15 structure.

\section{Boson Mott Insulator}

\textbf{Stability of Dianion X$^{2-}$ in K$_3$X.}
To substantiate our claim of stability of dianion X$^{2-}$, rather than trianion X$^{3-}$  we present some basic considerations of charge transfer in molecular solid K$_3$X. Situation is more complex than in ionic insulator NaCl, where electron affinity of Cl, ionization energy of Na and a Madelung sum solely determine energetics. Electron and lattice polarizability affects are small in NaCl. 

Situation in K$_3$X is different because of a complex molecular and metal atom environment \cite{EAffinityTheory}. In our estimate X molecule barely manages to capture two electrons. Our analysis is a useful first step, as we don’t know structures and real stoichiometry of many of the K$_3$X compounds experimentally.

We first focus on polycyclic aromatic hydrocarbons X. Consider a benzene molecule in free space. It has negative electron affinity  $\sim$ - 1.5 eV. That is, it can not bind an extra electron in free space. Naphthalene also has a negative electron affinity. Picene and p-terphenyl have some similarities. Picene has a small positive electron affinity E$_a \approx$ 0.5 eV in free space. If we add two electrons total binding energy becomes $ \approx - 2 E_a + \frac{e^2}{R} \approx $ + 3 eV. Here R is the linear dimensions of picene molecule. Thus two electron binding is not possible in free space; i.e., a dianion (picene)$^{2-}$ is not stable in free space, because of Coulomb barrier.

In the solid state environment various factors could help to overcome a charging energy $\sim$ 3 eV for transferring two electrons to the molecule X: i) electron correlation effects in the p-$\pi$ bonded system, that has been invoked to give stable molecular charge -2e singlets and positive pair binding energy, ii) Madelung energy gain, iii) electron, lattice, molecular orientational polarization energies, iv) cohesive energy in the Cooper pair Mott insulating dianion system arising from Josephson tunnelling. In our estimate these corrections do not add up to overcome charging energy of 3 eV per formula unit. 

Interestingly dication (K$_3)^{2+}$ can help overcome the large charging energy by contributing to overall cohesion of the solid from: i) energy gain by hybridization of the lone electron with K orbitals within the complex K$_3$ and with empty LUMO orbitals of neighbouring dianions and ii) metallic cohesion in the correlated metallic state of the renormalized metal atom band. 

In our estimate, with help from various cohesion mechanism one barely manages to have two electron transfer to the anion. Results from our simple consideration is consistent with theoretical calculations of references xx.  When we consider three electron transfer an enhanced charging energy overwhelms any of the contributions above and we do not form a stable solid containing transfer of all three valence electrons from K$_3$ to X.

Similar considerations suggests that K$_3$C$_{60}$ family we have stable dianions $(C_{60})^{2-}$. Simple fcc and A15 structures in this family makes this estimate more meaningful.

\textbf{Dianion X$^{2-}$ as a Cooper Pair Box.} Polycyclic aromatic hydrocarbon molecules support valence bond resonances. It is a correlated many electron state, a kind of \textit{molecular Mott insulator} made of $\pi$ electrons and well described as an RVB state a la' Pauling. This inbuilt correlation manifests in several ways: for example, as a large $\sim$ 2 eV splitting between the first excited triplet and singlet states in benzene.  One has to go beyond Huckel theory to capture this type of interesting effects arising from RVB physics in aromatic hydrocarbon molecules.

Two added electrons (instead of filling a single LUMO, as in Huckel theory) joins the $\pi$ electron pool, get entangled and becomes part of the valence bond resonance containing two extra electrons: i.e., (n-1) bond singlet pairs and two doublons. Here 2n is the number of carbon atoms (number of $\pi$ electrons) in the polycyclic hydrocarbon. This charge carrying state X$^{2-}$ is a soft state in the sense energy required to remove a Cooper pair is smaller than that in corresponding Huckel's theory. For p-terphenyl number of valence bonds is 9; consequently there are 9 neutral Cooper pairs in the box.

What we have is a Cooper pair box, where electron pairs, rather than single electron, can tunnel in and out easily, as we will see. Number of pre-existing Cooper pairs for p-terphenyl is 3 $\times$ 3 = 9

A good variational wave function for this is not a single Slater determinant but a RVB wave function, a Gutzwiller projected (correlated) state containing two extra electrons. For mathematical convenience and without loss of generality, we perform a particle hole transformation and write our state as if it contains two added holes (rather than two added electrons):
\be
|~{\rm{molecular~RVB}} \rangle = \prod_i (1-n_{i\uparrow}n_{i\downarrow})
(\sum_{ij} \phi_{ij} b^\dagger_{ij})^{n-1} |0\rangle
\ee
where $\phi$ is a suitable short range (Cooper) pair function. Doublons in this state are delocalized  within the molecule in a correlated fashion and allows for possibility for charge -2e pair to tunnel in and out.

Polaronic effects (known in poly para-phenylene (polymeric analogue of p-terphenyl) \cite{PolaronTheoryPterphenylene}), rather than electron correlation effects have been invoked for K$_3$(p-terphenyl) \cite{ptpSupCond}. Strong polaronic lattice distortions will make the polaron heavier by decreasing polaron hopping matrix element via Frank Condon overlap effects. Further, in short molecules, polaronic effects and lattice dimerization are less, because of quantum fluctuations arising from electron correlation effects; p-terphenyl is a three benzene ring system.

\textbf{Josephson Coupling and Boson Mott Insulator Formation.}
What is the nature of charge transfer between two molecules that are correlated electron systems ? 
Because of the large charging energy processes such as  X$^{2-}$ X$^{2-} \rightarrow $ X$^{1-}$ X$^{3-}$ (one electron transfer) or  X$^{2-}$ X$^{2-} \rightarrow $ X$^{0}$ X$^{4-}$ (two electron transfer) are highly improbable. We will not consider them. 

Physically meaningful electron transfer process are: $X^{2-}$ X$^{0} \rightarrow $ X$^{1-}$ X$^{1-}$ (one electron transfer) or  X$^{2-}$ X$^{0} \rightarrow $ X$^{0}$ X$^{2-}$ (two electron transfer).
We first focus on one electron tunnelling. At the level of Huckel theory this process is simple. It involves hopping matrix elements t$_m$ between two resonant LUMO orbitals of neighbouring molecules. Since we have a strongly correlated electron system we need to pull out an electron from a spin singlet charge -2e molecular RVB state. Electron transfer requires a breaking a singlet bond or a doublon in the correlated $\pi$ electron pool. Further, valence bond resonance in the receiving and accepting molecule also gets disturbed leading to a kind of electronic Frank Condon reduction (molecular wave function renormalization constant, called molecular Z \cite{AkbarGBMolecularZ}. This single electron transfer is strongly inhibited.

Let us consider two electron transfer, X$^{2-}$ X$^{0} \rightarrow $ X$^{0}$ X$^{2-}$. Pre-existing pairing correlations in both molecules will enhance this matrix element. In the case of two weakly superconductors this process is the Josephson tunneling \cite{Josephson} that takes advantage of the superconducting coherence present. What we have is only enhanced pairing correlations within each molecule. So we can not use Josephson or Ambagaokar Baratoff formula.  

We can use second order perturbation theory and get an estimate of pair tunneling matrix element t$_B \sim \frac{t_m^2}{E_c}$. Here E$_c$ approximately the singlet triplet gap of the molecule. 

An earlier work \cite{MihirPairTunnelling} on weakly coupled large U Hubbard clusters, studied in the context of interlayer pair tunnelling mechanism of Wheatley, Hsu and Anderson \cite{WHA.PairTunnelling} is relevant for our present discussion. This study shows that one electron hoppiing between two finite size Hubbard clusters enable two electron transfer and a consequent enhancement of pairing correlation within the molecules. Using this calculation we can also give a more accurate estimate of t$_B$.

Charging energy U$_{\rm B}$, the energy difference between two configurations X$^{2-}$ X$^{2-}$ and X$^{0}$ X$^{4-}$, is typically larger than Cooper pair band width $\sim$ 2zt$_{B}$ (here z is the number of neighbours.

Thus we have effectively a Boson Mott insulator with a finite Mott Hubbard gap. We call it Boson Mott insulator because low energy charge carrying excitations above the Mott Hubbard gap are charge $\pm$ 2e singlet states rather than spin-\half electron. The effective Hamiltonian of the Boson Mott insulator involves only charge -2e or Boson degree of freedom represented by operators b$_i$'s :

\be
H_{\rm B} = -t_{\rm B} \sum_{\langle ij \rangle} b^{\dagger}_{i} b^{}_{j} + H.c. + U_{B} \sum_{i} (b^{\dagger}_{i} b^{}_{i} - 1 )^2
\ee

A chemical potential fixes total number of charge -2e Bosons to be equal to total number of molecules.
This Boson Mott insulator is a potential high Tc superconductor, if only we can dope it. This superconductor will be special, as Cooper pair size is effectively size of X molecule.

There are theoretical evidences \cite{pTheoryValenti,pTheoryAoki,pTheoryYamada} that the stoichiometric compound K$_2$(aromatic molecule) are electron Mott insulators containing two electrons per X molecule. According to our theory these Mott insulators are best viewed as charge -2e Boson Mott insulators that also contain superconducting phase fluctuations. Our new insight leads to interesting consequences.

\section{Fermion Mott Insulator}|

\textbf{Renormalized Neutral Metal Atom and Mott Insulator Formation.} In our proposal charge -2e molecular (dianionic) matrix remains insulating. Each cation complex (K$_3)^{2+}$ carries one valence electron delocalized within the K$_3$ complex. Depending on the structure, not all three atoms are crystallographically equivalent and lone electron will be shared appropriately. 

What is important for us is that there is a lone electron residing in isolated cation complexes (K$_3)^{2+}$. These K atom states will hybridize through empty LUMO of neighbouring inert molecules and form a narrow renormalized metal atom band containing one electron per renormalized orbital. This is our renormalized metal atom Mott insulator described by the fermionic Hubbard model:

\be
H_{\rm F} = -t \sum_{\langle ij \sigma \rangle} c^{\dagger}_{i\sigma}c^{}_{j\sigma} + H.c. + U \sum_{i} n_{i\uparrow} n_{i\downarrow}
\ee

The band parameters t and U can be inferred from existing band structure calculations for picene and related systems. We will not present details now.

\textbf{Superconductivity via RVB Mechanism.} It is well recognized now, thanks to various development in RVB approach \cite{PWAScience,BZA} that spin-\half single orbital Mott insulators are seats of high Tc superconductivity for a range of doping. In real systems competing charge order, spin order and lattice distortion, when present, degrade superconductivity by localizing valence bonds, charges and spins. 
 
At the heart of RVB mechanism of superconductivity is use of valence bond or electron pairing correlation in spin-half Mott insulators. Doping of a Mott insulator is essential. An undoped neutral RVB state, though a coherent quantum fluid of spin paired electrons, is \textit{a charge incompressible state}. The Mott-Hubbard gap, in the context of an undoped spin-half Mott insulator (a half filled band) represents charge incompressibility. Doping produces a charge compressible charged RVB state, which in general is a superconducting state.

How do we get superconductivity without external doping in a half filled band ? Closeness of a half filled band of metal to Mott insulating state makes it a seat of superconductivity. Most organic superconductors begin as Mott insulators, with a band filling of half. Pressure causes an insulator to superconductor transition. What is important is that a metallic state  continues to support superexchange, when it is in the vicinity of a Mott insulating state. This is well described as a \textit{self-doped Mott insulating state} \cite{GBOrganics,McKenzee}, where, for energetic reasons, a Mott insulator spontaneously generates and maintains a small and equal density of holons (charge +e, spin-0) and doublons (charge -e, spin-0) in the background of superexchange and resonating singlets.

It is possible that many of the superconducting K$_3$X systems are at half filling and superconductivity arises because of its vicinity to a Mott insulating state. In our estimates using RVB theory we get superconducting Tc in the range of 100 K rather easily, for a reasonable choice of band parameters and structure. Since structures are not known precisely at the moment we will be contented with a ball park estimate.

In the next section we will see that presence of a Boson Mott insulator brings new possibilities through interesting interplay and synergy with the Fermionic Mott insulator.

\section{Superconductivity}

\textbf{Two Mott insulators and Prospects for Room Superconductivity.} We saw possibility of high Tc superconductivity in the Fermionic Mott insulator subsystem. In principle the Boson Mott insulator supported by X$^{2-}$ molecular subsystem, in view of a large pairing fluctuations, can encourage pairing in the fermion system through proximity effect. This will increase superconducting Tc to some extent.

There are two additional interesting possibilities. Electron affinities of the Fermion Mott insulator and Boson Mott insulator will be different in general. Beyond some threshold value there is possibility of transfer of electron pairs, for example from Boson Mott insulator to Fermion Mott insulator. Electron transfer is more likely, as adding charge -2e to X$^{2-}$ is more difficult than removing charge -2e. This internal charge transfer is a \textit{mutual doping}. In the present case Fermion Mott insulator is electron doped and Boson Mott insulator is hole doped. Several manybody effects contribute to charge transfer and determine density of charge transfer n. Two major factors contributing to them are superconducting condensation energy gain in both systems and Madelung energy gain arising from enhanced charge fluctuation.

Boson Mott insulator becomes a superconductor after transferring a density n of electron pairs. It is effectively a Bose condensation of holes in a lattice of hard core bosons carrying charge -2e. When hole density n is small, superconducting Tc of doped Boson Mott insulator is well approximated by Bose Einstein condensation formula:
\be
k_B Tc \approx 3.3125 ~ \frac{\hbar^2 n^{\frac{2}{3}}}{m^* k_{\rm B}}
\ee
We will not go into details, but just point out that effective mass of (hole) bosons (as estimated from pair tunnelling matrix element) and a reasonable hole boson density n gives us Tc's in the room temperature scales. This is an encouraging signal worth exploring further, looking for optimal m$^*$ and charge transfer n.

Fermion Mott insulator which becomes superconducting via RVB mechanism will have its own Tc, determined by different parameters. Coupled order parameters from two superconductors makes the problem richer.

The synergy is interesting. Boson Mott insulator gives electrons, Fermion Mott insulator gives holes and both become superconducting. We call this \textit{shared superconductivity}. 

Second interesting possibility is existence of a compound K$_{2-x}$X, with a stoichiometry close to two. In this case fermion Mott insulator is not present. However, we have a hole doped Boson Mott insulator. This system also can lead to high Tc superconductivity.

Since we have several candidate aromatic hydrocarbon and C$_{60}$ systems, parameter space is big and we could realize some of these interesting possibilities. 

\textbf{Two Mott Insulator Theory of Superconductivity in C$_{60}$ based Systems.} In the superconducting compound K$_3$C$_{60}$, fullerene molecules form a fcc lattice. Fcc lattice has octahedral and tetrahedral interstitial sites in the ratio 1:2. Alkali atoms fill in all interstitial sites. Octahedral interstitial sites also form an fcc lattice. Tetrahedral interstitial sites form a simple cubic lattice, which is symmetrically inscribed in the fcc lattice of octahedral interstitials. As interstitial lattices have different geometrical environment, distance between K atom and its nearest C$_{60}$ molecule is closer for tetrahedral sites than octahedral sites.

According to our hypothesis, two K atoms in tetrahedral interstitials transfer two electrons to one C$_{60}$ molecule. K atom from octahedral interstitial, one per formula unit, retains a valence electron. Thus nominal charge state of our reference system is K$^0 (K_2)^{2+}$ C$_{60}^{2-}$. 

This is partly because of closeness of K atoms at tetrahedral interstitial to four C$_{60}$ molecules. Stability of molecular singlet or a positive pair binding energy and dynamical Jahn Teller distortion also help two electron transfer and creation of a stable charge -2e spin singlet molecular state. Further Madelung energy gain decreases because of dielectric screening, making charge three electron transfer less likely. It is interesting to note that in a recent model calculation for pair binding in C$_{60}$, using DMRG method, an orbitally symmetric and spin singlet stable state with maximum pair binding has been found.

According to our hypothesis single valence electron of K atom in octahedral interstitial hybridize with LUMO of C$_{60}^{2-}$ molecule and forms a single orbital half filled correlated metallic state in an fcc lattice. This metal is close to the Mott insulator boundary and support RVB mechanism based superconductivity. The well known experimental fact \cite{IwasaC60} that intercalation with NH$_3$ expands the lattice and converts NH$_3K_3$C$_{60}$ to a Mott insulator is not in contradiction with our hypothesis.

It is also interesting that highest Tc in the alkali fulleride family \cite{RbCs2C60} is exhibited by RbCs$_2$C$_{60}$. This fits with our hypothesis as well, if Cs occupies the tetrahedral site completely.  Further, Cs has lower ionization energy compared to Rb and hence transfers two of its electron and makes a Bosonic Mott insulator in the dianion molecular matrix C$_{60}^{2-}$

Mott insulating Cs$_3C_{60}$ in the A15 structure is interesting \cite{A15C60} from the point of view of our hypothesis. Fullerene molecules form a bcc lattice. Cs atoms fill in interstitial sites such that pairs of Cs atoms lie on three sets of parallel faces in mutually perpendicular direction. We get three sets of Cs chains. There are two possibilities. 

First one is, a stronger version of our hypothesis - there is no charge transfer from Cs to the C$_{60}$ sub system at all. We have essentially created an expanded lattice of Cs atoms, where Mott physics begins to play an important role and creates a spin-\half renormalized Cs atom Mott insulator. Absence of charge transfer from K atoms in this system is a distinct possibility, because of an increase in distance between K atom and its nearest fullerene molecule in this A15 structure. Pressure can induce a Mott insulator to metal transition and lead to superconductivity via RVB mechanism in the vicinity of the insulator to metal boundary.

Second intriguing possibility is creation of a Boson Mott insulator as before, C$_{60}^{2-}$, by each fullerene molecule accepting just two electrons. Remaining single valence electron in the cation complex K$(K_2)^{2+}$ forms a Mott insulator; it can happen in few different ways. One such possibility is that the three sets of K atom chains alluded to earlier have a rational occupancy of $\frac{1}{3}$ and forms a Wigner-Mott lattice at ambient pressure. Wigner-Mott lattice undergo metallization under pressure leading to superconductivity. In this case there is a lattice symmetry breaking misfit between symmetry of the K atom subsystem and C$_{60}$ subsystem. It is also possible that mutual doping, discussed earlier, creates superconductivity in both Mott insulators.

\textbf{Superconductivity in K$_3$(p-terphenyl).}
Stoichiometric K$_3$(p-terphenyl), according to our theory, has a Boson and a Fermion Mott insulator. 
In the absence of detailed crystal structure and other phenomenological inputs it is difficult to say what is the origin of superconductivity in this fascinating system, within the two Mott insulator model. We wish to summarize the three possibilities we have spelled out before: i) a charge transfer from Boson to Fermion Mott insulator (mutual doping) and both exhibiting superconductivity, ii) Fermion part is superconducting, as it is a half filled band metal but close to the Mott transition point; Boson part remains essentially an insulator and iii) stoichiometry is close to K$_{2-x}$X and we have a hole doped Boson Mott insulator, which exhibits superconductivity.

Existing phenomenology, namely a superconducting gap of about 20 to 25 meV seen in high resolution PES \cite{ptpPES} and relatively smaller H$_{c1}$ \cite{ptpSupCond} implies that doped Boson Mott insulator is playing some key role in establishing superconductivity \cite{PWAprivateCom}. Pairing energy scale (pair binding energy) within the molecule (p-terphenyl)$^{2-}$ is high. It is likely that gap seen in the experiment is from induced (proximity) superconductivity in the Fermionic Mott insulator.

As we mentioned earlier, we do not find serious theoretical constraints on reaching higher Tc's in these family of compounds. 

\section{Discussion}

In this article we have suggested a novel possibility of formation of a Boson and fermion Mott insulator in the alkali doped aromatic hydrocarbon and fulleride family K$_3$X. We have discussed superconductivity in this correlated electron system, from RVB mechanism point of view. It is remarkable that RVB physics works in two ways: i) RVB physics inside the molecule X gives rise to a Boson Mott insulator and ii) Fermionic Mott insulator of the metal atom  leads to the standard RVB physics involving valence bonds living in the lattice of the metal atom complexes $(K_3)^{2-}$.

Metal ammonia solution \cite{Ogg,MetalAmmoniaSolnReview} has been some what of a mystery in condensed matter community, ever since Ogg's observation of Meissner signals at room temperature in 1946. We find that the two Mott insulator scenario we have suggested for K$_3$X has an interesting role to play \cite{GBUnpublished} for metal ammonia solid solutions and related systems.

\textbf{Acknowledgement}. I thank V Awana (NPL, New Delhi) for alerting me about reference \cite{ptpSupCond}. It is a pleasure to thank P.W. Anderson for discussions. I thank SERB, Government of India, for a Distinguished Fellowship. This work, partly performed at the Perimeter Institute for Theoretical Physics, Waterloo, Canada is supported by the Government of Canada through Industry Canada and by the Province of Ontario through the Ministry of Research and Innovation.


\begin{thebibliography}{99}
\bibitem{picenePterphenyl} Picene, C$_{22}$H$_{10}$, a pioneer in the aromatic hydrocarbon series of superconductors, is structurally very close to p-terphenyl molecule. Picene is obtained by removing four H atoms in p-terphenyl and replacing them by a pair of bonded C-C atoms, in suitable places.
\end{thebibliography}
\end{document}